\documentstyle[11pt,epsfig]{article}
\newcommand \be{\begin{equation}}
\newcommand \ba{\begin{eqnarray}}
\newcommand \ee{\end{equation}}
\newcommand \ea{\end{eqnarray}}

\begin{document}

\begin{center}
{\LARGE Towards Landslide Predictions: Two Case Studies}
\end{center}
\bigskip
\begin{center}
{\large D. Sornette$^{1-3}$, A. Helmstetter$^{3}$,
J. V. Andersen$^{1,5}$, S. Gluzman$^3$, J.-R. Grasso$^{3,4}$ and V. 
Pisarenko$^6$}
\end{center}
\bigskip
\begin{center}
{$^1$ LPMC, CNRS UMR 6622 and
Universit\'{e} de Nice-Sophia Antipolis\\ Parc Valrose, 06108 Nice, France\\
$^2$ Department of Earth and Space Sciences\\
   University of California, Los Angeles, California 90095-1567\\
$^3$ Institute of Geophysics and Planetary Physics\\
University of California, Los Angeles, California 90095-1567\\
$^4$ LGIT, Observatoire de Grenoble, Universit\'e Joseph Fourier, France\\
$^5$ U. F. R. de Sciences Economiques, Gestion, Math\'ematiques et
Informatique, \\ CNRS UMR7536 and Universit\'e Paris X-Nanterre \\
92001 Nanterre Cedex, France\\
$^6$ International Institute of Earthquake Prediction Theory and
Mathematical Geophysics\\ Russian Ac. Sci. Warshavskoye sh., 79, kor. 2,
Moscow 113556, Russia}
\end{center}

\begin{abstract}

In a previous work \cite{comphelmjgr}, we have proposed a simple physical model
to explain the accelerating displacements preceding some catastrophic 
landslides,
based on a slider-block model with a state and velocity dependent friction law.
This model predicts two regimes of sliding, stable and unstable leading to
a critical finite-time singularity.
This model was calibrated quantitatively to the displacement
and velocity data preceding two landslides, Vaiont (Italian Alps) and 
La Clapi\`ere
(French Alps), showing that the former (resp. later) landslide is
in the unstable (resp. stable) sliding regime. Here, we test the predictive
skills of the state-and-velocity-dependent model on these two landslides, using
a variety of techniques.
For the Vaiont landslide, our model provides good predictions of
the critical time of failure up to 20 days before the collapse.
Tests are also presented on the predictability of the time of the 
change of regime for
la Clapi\`ere landslide.

\end{abstract}

\section{Introduction}

There is a growing interest in understanding and predicting catastrophic
phenomena, such as floods, earthquakes and avalanches, which are
characterized by their rareness and burstiness often leading to disastrous
consequences for the embedding environment \cite{BKS}. Notwithstanding their
large societal impacts, the scientific community is only beginning
to develop the concepts and tools to model and predict these class of events.
The prediction of
catastrophes is often considered to be essentially impossible either due to
intrinsic mechanisms inherent to the systems \cite{Gellerjackson}
or from the existence of enormous
practical barriers \cite{karplus}. Several groups have however
found evidence of a degree of predictability of certain catastrophes 
\cite{SorPNAS},
such as financial crashes \cite{sorbookcrash} and earthquakes 
\cite{Keilisbook}.

Here, we address the
question of the predictability of landslides
which constitute a major geologic hazard of strong concern in most
parts of the world.
Landslides occur in a wide variety of geomechanical contexts,
geological and structural settings, and as a response to
various loading and triggering processes. They are often associated with
other major natural disasters such as
earthquakes, floods and volcanic eruptions.

Derived from the civil-engineering methods developed for
the safety of human-built structures, including dams and bridges, the
standard approach to slope instability is to identify
the conditions under which a slope becomes unstable \cite{Hoek2}.
By their nature, standard stability analysis cannot account for
acceleration in slope movement \cite{Hoek1}. The problem is
that this modeling strategy gives a nothing-or-all signal.
In this view, any specific landslide is essentially unpredictable,
and the focus is on the recognition of landslide prone areas.
This approach is very similar to the practice in seismology called
``time-independent hazard'' where earthquake prone areas are located
in association with active faults for instance, while the prediction of
individual earthquake is recognized to be much more difficult
if not unattainable. This ``time-independent hazard'' essentially
amounts to assume that landslides are a random
(Poisson)  process in time, and uses geomechanical modeling to
constrain the future long-term landslide hazard.
The approaches in terms of a safety factor do not address the
preparatory stage leading to the catastrophic collapse, if any.
In contrast,``time-dependent hazard'' would accept a degree of
predictability in the process, in that the landslide hazard varies with time,
maybe in association with varying external forcing (rain, snow, 
earthquake, volcano).
The next level in the hierarchy would be ``landslide forecasting'',
which require significant better understanding to allow for the
prediction of some of the features of an impending landslide,
usually on the basis of the observation of precursory signals. Practical
difficulties include identifying and measuring reliable, unambiguous
precursors, and the acceptance of an inherent proportion of missed
events or false alarms.

Our purpose is to extend the model of a slider-block with 
state-and-velocity-dependent
friction introduced in our companion paper \cite{comphelmjgr}
for the analysis of two landslides.
Here, we examine how one could have perhaps predicted these 
landslides in advance.
By studying these two cases, we hope
to develop a methodology that could be useful in the future as well as
to determine the limits of predictability.
For the Vaiont landslide, our model provides good predictions with a precision
of about one day of
the critical time of failure up to 20 days before the collapse.
Tests are also presented on the prediction of the time of the change 
of regime for
la Clapi\`ere landslide.
Re-examining the calibration of the model of a slider-block with 
state-and-velocity-dependent
friction, we cannot
exclude that La Clapi\`ere might also belong to the unstable velocity weakening
regime; its deceleration observed after 1988 may then be interpreted 
as a change of surface
properties that modifies the friction law parameters.

\section{A short synthesis of time-dependent predictive approaches}

\subsection{Phenomenological power law acceleration}

Accelerating displacements preceding some catastrophic landslides have been
found empirically to follow a time-to-failure power law, corresponding
to a finite-time singularity of the
velocity $v \sim 1/(t_c-t)$ \cite{Voight88}.
Controlled experiments on landslides driven by a monotonic
load increase have been quantified by a scaling law relating
the surface acceleration $d\dot{\delta}/dt$
to the surface velocity $\dot{\delta}$ according to
\be
d\dot{\delta}/dt = A \dot{\delta}^{\alpha}~,
\label{mgnlsa}
\ee
where $A$ and ${\alpha}$ are
empirical constants \cite{Fuko}. For ${\alpha}>1$, this relationship
predicts a divergence of the sliding velocity in finite time at some critical
time $t_c$. The divergence is of course not to be taken literally: it signals
a bifurcation from accelerated creep to complete slope instability
for which inertia is no more negligible. Several cases have been
quantified ex-post with this law, usually for ${\alpha}=2$, by plotting the
time $t_c-t$ to failure as a function of the inverse of the creep velocity
(see \cite{Bhandari} for a review). Indeed, integrating (\ref{mgnlsa}) gives
\be
t_c-t \sim \left({1 \over \dot{\delta}}\right)^{1 \over \alpha-1}~.
\label{jmgjkla}
\ee
For the Mont Toc, Vaiont landslide revisited here,
Voight mentioned that a prediction of the failure date
could have been made more than 10 days before the actual failure,
by using a linear relation linking the inverse velocity and the time
to failure, as found from (\ref{jmgjkla}) for $\alpha=2$ \cite{Voight88}.
Our goal will
be to avoid such an a priori postulate by calibrating a more general
physically-based model for the purpose of forecasting.

\subsection{Slider-Block model with state and velocity dependent friction}

We briefly summarize Ref.~\cite{comphelmjgr}, which models the future
landslide as a block resting on an inclined slope forming a fixed angle
$\phi$ with respect to the horizontal.
The solid friction coefficient $\mu$ between two surfaces
is taken to be a function of the cumulative slip $\delta$ and the 
slip velocity $\dot{\delta}$
according to the Dieterich-Ruina law \cite{Diete78,Ruina}:
\be
\mu = \mu_0 + A \ln {\dot{\delta} \over \dot{\delta}_0} + B \ln
{\theta \over \theta_0}~,
\label{vxcxxzx}
\ee
where the state variable $\theta$ is usually interpreted as
proportional to the surface
of contact between asperities of the two surfaces. $\mu_0$ is the
friction coefficient for a sliding velocity $\dot{\delta}_0$
and a state variable $\theta_0$.
The state variable $\theta$  evolves with time according to
\be
{{\rm d}\theta \over {\rm d}t} = 1 - {\theta \dot{\delta} \over
D_c}~, \label{qwertyu}
\ee
where $D_c$ is a characteristic slip distance, usually interpreted as
the typical size of asperities. Expression (\ref{qwertyu})
can be rewritten as
\be
{{\rm d}\theta \over {\rm d}\delta} = {1 \over \dot{\delta}}
- {\theta \over D_c}~. \label{qweadasgatyu}
\ee

For $m \neq 1$, it is convenient to introduce the reduced variables
\be
x \equiv (S \theta_0)^{1/(1-m)}~{\theta \over \theta_0}~,
\label{mgmbm}
\ee
and
\be
D \equiv D_c~\left(S \theta_0^m \right)^{1 \over 1-m}~,
\label{hfff}
\ee
where
\be
S \equiv {\dot{\delta}_0 ~ e^{{\tau \over \sigma}-\mu_0 \over A} \over D_c}
\label{mmjdl}
\ee
and $m = B/A$. $\tau$ and $\sigma$ are the average shear and normal
stresses at the sliding interface of the block.
Then, expression (\ref{vxcxxzx}) reads
\be
{\dot{\delta} \over \dot{\delta}_0} = D ~x^{-m}~.
\label{velocity}
\ee
Similarly, expression (\ref{qwertyu}) transforms into
\be
{{\rm d}x \over {\rm d}t'} = 1 - x^{1-m}~,
\label{ode}
\ee
where $t' = t/T$ with
\be
T = {D_c \over D} =
\left[{D_c \over \dot{\delta}_0 \theta_0^m}\right]^{1/(1-m)}~
e^{{\tau \over \sigma}-\mu_0 \over B-A}~.
\label{mgmlla}
\ee
In the sequel, we shall drop the prime and use the dimensionless
time $t'$, meaning that time is expressed in units of $T$ except
stated otherwise.
The case $m=1$ requires a special treatment \cite{comphelmjgr}.

In our previous
companion paper \cite{comphelmjgr}, we have analyzed this set of
equations (\ref{velocity},\ref{ode}) and shown that it provides a
physical basis for the phenomenological law (\ref{jmgjkla}).
Indeed, it is easy to show that, for $m>1$ and $x_i<1$ and
sufficiently close to the singularity $t_c$, the
slip velocity is of the form (\ref{jmgjkla}) with $\alpha = 2$,
that is, the slip velocity is inversely proportional to time.
Consequently, the slip $\delta(t) \sim \ln [1/(t_c-t)]$ diverges
logarithmically.

More generally, depending on the ratio $m=B/A$ of
two parameters of the rate and state friction law and on the initial
frictional state of the sliding surfaces characterized by the reduced
parameter $x_i=x(t=0)$ defined in (\ref{mgmbm}), four possible 
regimes are found.
Two regimes can account for an acceleration of the displacement.
For $B/A>1$ (velocity weakening) and $x_i<1$,
the slider block exhibits an unstable acceleration
leading to a finite-time singularity of the displacement
and of the velocity $\dot{\delta} \sim 1/(t_c-t)$, thus rationalizing 
Voight's empirical law.
An acceleration of the displacement can also be reproduced in the
velocity  strengthening regime, for $B/A<1$ and $x_i>1$. In this case,
the acceleration of the displacement evolves toward a stable sliding with
a constant sliding velocity.
The two others cases ($B/A<1$ and $x_i<1$, and $B/A>1$ and $x_i>1$) give
a deceleration of the displacement.
We have used the slider-block friction model to analyze 
quantitatively the displacement
and velocity data preceding two landslides, Vaiont (in the Italian 
Alps) and La Clapi\`ere
(in the French Alps) \cite{comphelmjgr}.
The Vaiont landslide was the catastrophic culmination of an accelerated slope
velocity. La Clapi\`ere landslide was characterized by a strong slope
acceleration over a two years period, succeeded by a restabilizing phase.
Our inversion of the slider-block model on these
data sets showed good fits and suggested to classify the Vaiont
(respectively La Clapi\`ere) landslide as belonging to the velocity weakening
unstable (respectively strengthening stable) sliding regime.

\section{Prediction of the Vaiont landslide}

On October 9, 1963, a 2 km-wide landslide initiating at an elevation of
1100-1200 m, that is 500-600 m above the valley floor,
on the Mt Toc slope in the Dolomite region in the Italian Alps
about 100 km north of Venice, ended up 70 days later in a 20 m/s run-away of
about 0.3 km$^3$ of rocks sliding into a dam reservoir.
The high velocity of the slide triggered  a water surge within the reservoir,
overtopping the dam and killing 2000 people in the village downstream.
For a synthesis of its history and the analysis of time series of
slip velocity of benchmarks on its flanks, we refer to our companion paper
\cite{comphelmjgr}.

\subsection{Analysis of the cumulative displacement data with the
slider-block model parameters}

Previously, we have calibrated the slider-block model with 
state-and-velocity-dependent
friction on the time series of slip velocities of several benchmarks. The
key parameter $m=B/A$ is found to be larger than $1$, indicating
an unstable regime leading to a finite-time singularity \cite{comphelmjgr}.
However, we noted that
the parameters of the friction law are poorly constrained by the inversion.
In particular, even for those benchmarks with the best fit gives $m>1$,
other models with $m<1$ provide a good fit to the velocity with only slightly
larger residuals.

\begin{figure}
\epsfig{file=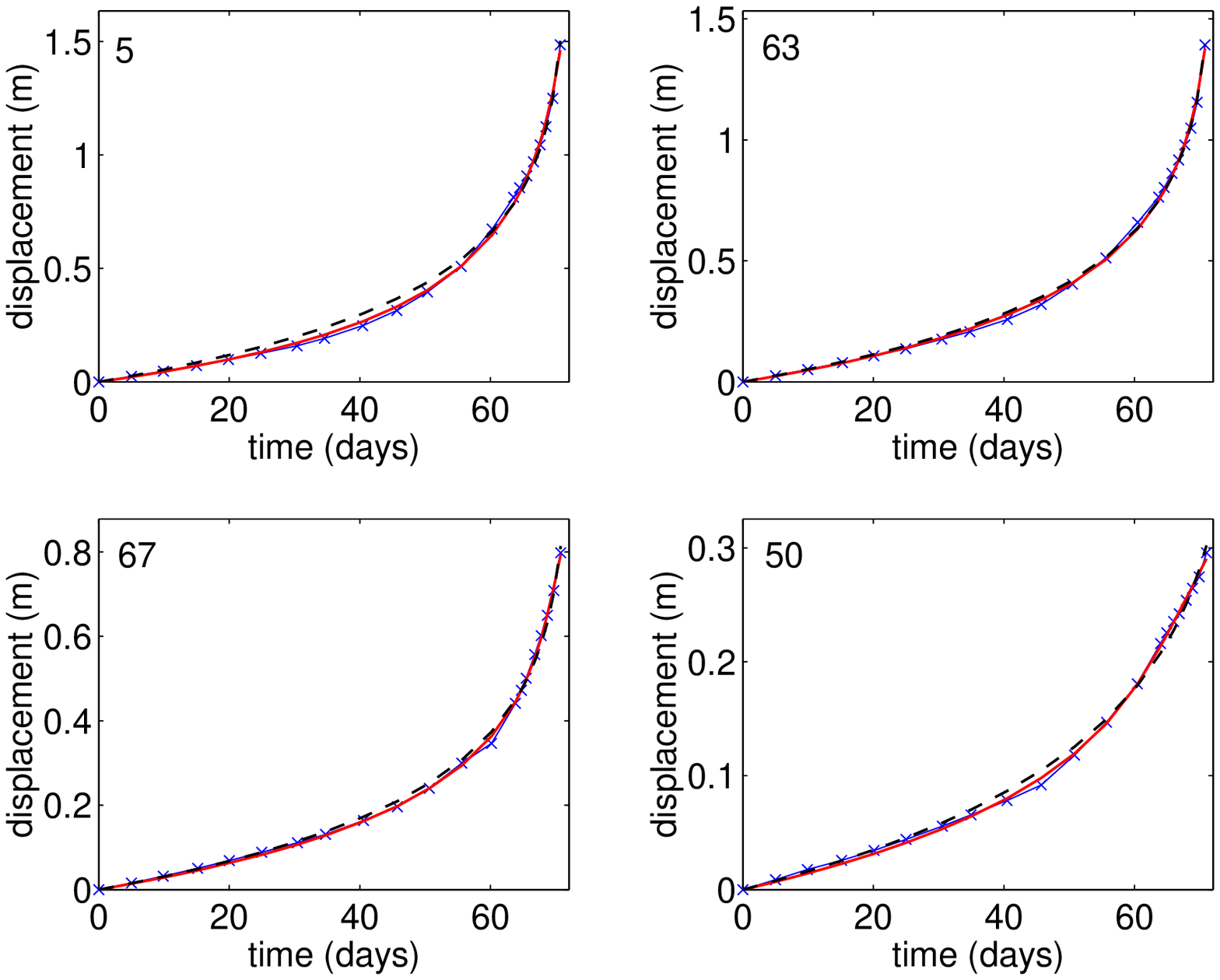,width=14cm}
\caption{\label{depvaiont} For each of the four Vaiont benchmarks,
the cumulative displacement data
is fitted with the slider-block model with the state and velocity
friction law (\ref{ode}) and (\ref{velocity}) by adjusting the set of
parameters $m$, $D/T$ and the initial condition of the state variable $x_i$.
The data is shown as the crosses linked by straight segments and the fit
is the thin continuous line. The fitted $m$ are respectively
$m=0.99$ (benchmark 5), $m=0.85$ (benchmark 63), $m=0.68$ (benchmark 67)
and $m=0.17$ (benchmark 50). The fits with the slider-block model obtained
by imposing the value $m=1.5$ are shown with the dashed line for comparison.}
\end{figure}

We now contrast these results with those obtained by fitting
the cumulative displacement (rather than the velocity) with the slider-block
model with the state and velocity friction law (\ref{velocity})
and (\ref{ode}).
The results are shown in Figure \ref{depvaiont}.
The fitted $m$ are respectively $m=0.99$ (benchmark 5), $m=0.85$ 
(benchmark 63),
$m=0.68$ (benchmark 67) and $m=0.17$ (benchmark 50). These values differ
significantly from those obtained by the inversion of the velocity data and,
to make things worse, they all correspond to the velocity-strengthening
regime $m<1$.
At first sight, these results are quite surprising
since we fit the same data, the only difference being that the cumulative
displacement is the integral of the velocity. We think that the reason
for these discrepancies lies in the fact that, assuming that the
velocity-weakening regime $m>1$ holds, the corresponding logarithmic
dependence $\delta(t) \sim \ln(1/t_c-t)$ of the displacement $\delta$ 
is extremely
degenerate in that it predicts an acceleration of the displacement
which is significant only very close to the critical time $t_c$.
Therefore, a cross-over from a low velocity to a larger velocity
described by the regime $m<1$ may be selected by the inversion, as we
witness here.
This is a rather standard problem of logarithmic singularities,
which are so weak at providing
constraints, notwithstanding the a priori reduction of noise
obtained by constructing a cumulative quantity. It may actually be the case
that the cumulative noise deriving from the integral of the velocity
is enough to spoil the weak logarithmic singularity 
\cite{Huangetal00}: the resulting
correlated noise seems to select a milder behavior.
We are thus led to conclude that fits to the sliding velocity which involves
stronger power law singularities should be more reliable and we shall use
them exclusively in our prediction tests reported for the Vaiont landslide.

\subsection{Initiation of the instability and tests of robustness}

The critical acceleration is observed neatly only in the last 70
days before the catastrophic landslide. Before,
the alternation of phases of accelerating and decelerating velocity
in the 1960-1962 period implies that some friction parameters have
changed, maybe due to changes in water level, resulting in a change
of sliding regime.
The change of water level may have modified the material properties
of the underlying solid contacts at the base of the moving rock mass
\cite{Erismann}, therefore changing the parameter
$m=B/A$ from the stable to the unstable regime.
Another possibility is that changes in water level have
modified the population of contacts at the basis of the rock mass,
therefore changing the parameters of the friction law,
and changing the sliding regime from the decelerating regime
to to the accelerating regime. One possible simple change
of the parameters of the friction law correspond to a change of
the initial condition on the state variable $x_i$, which may induce a change
of the sliding regime from the decelerating regime for $m>1$
and $x_i>1$ to the accelerating regime for  $m>1$ and $x_i<1$ and vice-versa.

We have also tried to invert the friction law parameters
using only data up to a time $t_{max}$ smaller than the last available point
(equal to 70 days from the origin of the time series)
before the catastrophic landslide occurs,
to mimic a real-time situation.
Changing $t_{max}$ between 30 and 70 days, we obtain a large variability of
the parameters.
Most values $m$ are found larger than 1 for $30<t_{max}<55$ days,
and then become smaller than one, and return to $m \geq 1$ for 3 
benchmarks when
using the full velocity data.
Similar fluctuations are found when using a synthetic data set generated with
the friction model. We have generated a synthetic data set using the same
parameters as those of the best fit of benchmark 5,
and added a white noise
with the same standard deviation as that of the residue of the fit of 
benchmark 5.
Although this synthetic data set was generated with $m=1.35$, both
$m>1$ and $m<1$ (for 2 points over 15 points) values are obtained 
when inverting
the parameters up to $t_{max}$ and changing $t_{max}$ between 30 and 70 days.
However, values with $m<1$ for this synthetic data set are much less 
frequent than
for the Vaiont velocity data in relative terms.

\subsection{Predictions and ex-post skills}

We present a series of attempts at predicting in advance the critical
time $t_c$ of the catastrophic Vaiont landslide instability. These
attempts rely solely on the analysis of the four benchmarks velocity
data up to various times $t_{\rm max} < t_c$ mimicking a real-time situation.
Therefore, we truncate the data at some time $t_{\rm max} < t_c$ and use
only the data up to $t_{\rm max}$. Our goal is i) to investigate whether
a prediction in advance could have been issued, as suggested by 
Voight \cite{Voight88},
ii) to establish the reliability and the precision limits of such 
predictions and iii)
to test various prediction schemes that we have developed in the 
recent past for other
applications or specifically for this problem.
We use and compare three methods to predict the critical time 
$t_c=69$ days of the collapse
\begin{itemize}
\item the slider block model with the state and velocity friction law 
described above;
\item an approximation of the slider block model based on the functional
renormalization method described below;
\item a simple finite-time singularity (\ref{jmgjkla}) with $\alpha=2$ as
proposed by Voight \cite{Voight88}.
\end{itemize}

\subsubsection{Prediction using the slider-block model with the state
and velocity friction law}

The prediction of the critical time $t_c$ is obtained by fitting the
slider-block model on the velocity time series of the four benchmarks up to
a time  $t_{\rm max}$. For $m \geq 1$, $t_c$ is the time of the
divergence. The divergence of the velocity exists only in the unstable
regime $m>1$. Therefore, we choose the best fit with $m>1$, even if the
best model gives sometimes $m<1$.

\subsubsection{Functional renormalization of the friction law}

We are dealing with a noisy time series with relatively few data points
for which the detection of a singularity is a difficult task. Rather than
using the full solution of a model assumed to be a good representation
of reality as done in the previous sections, it may be profitable to develop
prediction schemes that are less constrained by the necessarily
restricting physical assumptions underlying the model
and that are more specifically designed
from a mathematical point of view to be resilient to
noise and to the scarcity of data. Such a method is the
so-called functional renormalization method,
which constructs the extrapolation for future time $t>t_{\rm max}$
from a re-summation of the time series represented by a simple polynomial
expansion in powers of time $t$. Its mathematical foundation has been developed
in a series of papers 
\cite{YukalovGluzman97,GluzmanYukalov97,GluzmanYukalov98}.
The application of this method to detect and predict finite-time singularities
has been already investigated in \cite{Gluzmanetal01,GluzmanSornette02}.
We refer to these papers for a presentation
of the method and restrict ourselves here to the concrete application
of the method to the friction law  (\ref{velocity}) and  (\ref{ode}).

The first input of the functional renormalization approach is an expansion
of the variable to be predicted in increasing powers of time.
In our case, we use the  functional renormalization approach to
provide an approximate analytical solution of the differential equation
of the friction model (\ref{ode}).
This method is much more efficient numerically than the numerical
resolution of the differential equation  (\ref{ode}).
The friction model (\ref{ode}) gives the time evolution of the
state variable from which the sliding velocity $\dot{\delta}$ derives
using (\ref{velocity}).

The needed expansion of $y\equiv \theta/\theta_0$ in powers of time $t$
is obtained from a Taylor expansion whose coefficients are derived from
successive differentiation of (\ref{ode}). Up to fourth order $t^4$,
calling $y_0 = \theta(t=0) /\theta_0$, we obtain
\begin{equation}
\label{6}
y_k(t) \simeq \sum_{n=0}^k a_n t^n,~~~~~ t\rightarrow 0,~~~~~~k=1, 2, 3, 4,
\end{equation}
where the coefficients $a_n$ are given in the Appendix A as a function of the
friction parameters and of the initial condition.

The functional renormalization approach is in principle able to derive
an extrapolation to the future from the form (\ref{6}). However, in
order to obtain an optimal stabilization, it is essential to incorporate
as much available information as possible. In particular, in our case,
we know the functional form of the dependence of the state variable as a
function of time in the asymptotic regime (large times for $m<1$ and close
to the singularity for $m>1$). Therefore, the second input of our
implementation of the functional renormalization approach is the following.
For $m<1$, in a long-time limit, it is easy to show that equation
(\ref{ode}) has an asymptotic solution in the form,
\begin{equation}
\label{10}
y_{t \to \infty}(t)\simeq y^{*}+A_1\exp \left( -\frac t{t^{*}}\right) +A_2\exp
\left( -\frac{2t}{t^{*}}\right) + {\rm h.o.t.}
\end{equation}
where $1/t^* = (1-m)/T = (1-m) (S\theta_0)^{1/(1-m)}$
and h.o.t. stands for higher-order terms.
The coefficients $A_1$and $A_2$ are unspecified at this stage and
can be determined using the crossover technique \cite{GluzmanYukalov98},
in order to optimize the stability of the solution.
For $m\geq 1$, the asymptotic expression as $t \rightarrow t_c$ is of the
form
\be
x(t) \simeq  m^{1 \over m} ~(t_c-t)^{1 \over m} ~,
\label{mmsls}
\ee
where the critical time $t_c$ is given by expression (\ref{23}).
However, we shall allow the prefactor and $t_c$ to be adjusted
to ensure maximum stability. Specifically, the determined value of
$t_c$ will be a primary result of the crossover technique.

Our goal is thus to construct a function $y(t)$
which incorporates the short and long time asymptotics of the solution
as given by expressions (\ref{6}) and (\ref{10}) for $m<1$
and by (\ref{6}) and (\ref{mmsls}) for $m \geq 1$, while
possibly departing from it at intermediate times to allow for a
maximum stability.
The general mathematical formulas that are the solution of this 
problem are given
in Appendix A for the two cases $m<1$ and for $m \geq 1$ respectively.

For the application to the Vaiont landslide, and for each
``present time'' $t_{\rm max}$, we assume that $m>1$ so that $t_c$ exists
and we fit the expression of the fourth-order
approximate $y_4^{*}(t)$  given by (\ref{22}) to the velocity of each 
of the four
benchmarks, extract the corresponding parameters and put them in
equation (\ref{23}) in Appendix A for the critical time $t_{c4}$.
We stress that the function thus reconstructed is essentially
indistinguishable from the fit with the slider-block
friction model. Solving (\ref{23}) for $t_{c4}$ allows us to construct
the predicted critical time as a function of the ``present time'' 
$t_{\rm max}$.
We also estimate the value of $m$ as a function of  $t_{\rm max}$.
Apart from some large jumps that may be attributed to the sensitivity
of specific noisy points as $t_{\rm max}$ is scanned, we observe that
most fits are compatible with a value of $m$ in the range $1.3-1.5$.

\subsubsection{Finite-time singularity (\ref{jmgjkla}) with $\alpha=2$}

We use a simple linear regression of the inverse of the velocity as a
function of time, as proposed by Voight \cite{Voight88}.
We have found that, in order to have more stable parameters, it is
necessary to give less weight to the
early times where the velocity is small and contains little information
on the critical time. We find that weighting each data
point proportionally to its velocity provides stable fits.
The critical time $t_c$ is then given as the time at which the fitted
straight line of the $1/\dot{\delta}$ data intersects with the time axis.
Recall that a linear relation between $1/\dot{\delta}$ and time $t$ 
is equivalent
to a power law singularity of the velocity $\dot{\delta} \sim 1/(t_c-t)$,
as discussed previously, which is expected asymptotically close
to $t_c$ for the friction model in the case $m>1$ and $x_i<1$.

\subsubsection{Comparison of three different methods of prediction of $t_c$
as a function of the ``present time'' $t_{\rm max}$}

\begin{figure}
\epsfig{file=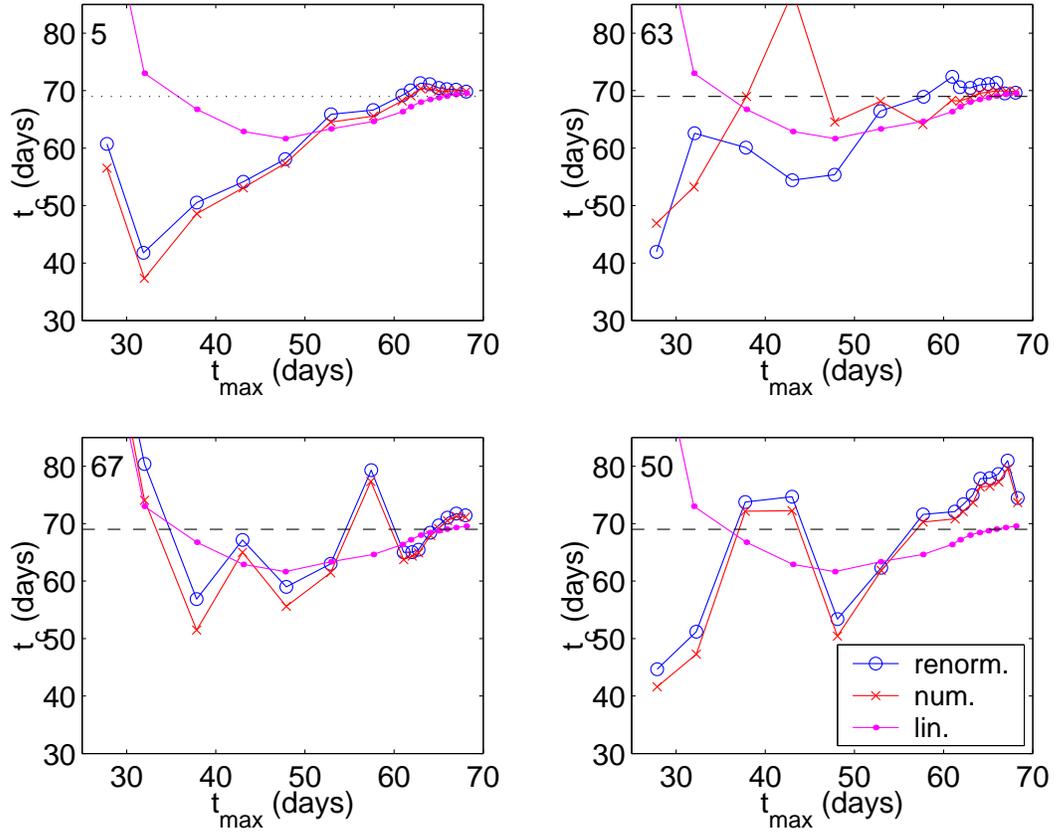,width=14cm}
\caption{Predicted critical time $t_c$ as a function
of the ``present time'' $t_{max}$ (last point used for the fit) for
all four benchmarks of the Vaiont landslide, using three
different methods of prediction described in the text:
renormalization method (circles), numerical evaluation of the
friction model (\ref{ode}) (crosses), and  linear regression of the inverse
velocity as a function of time performed by
removing the first point (early time) of the curve and using a weight
proportional to the velocity (dots). The horizontal
dashed line indicated the true critical time $t_c=69.5$ days  (for an
arbitrary origin of time from which the fits are performed to the
catastrophic landslide.
All methods impose $m>1$, but in some cases a better fit may be obtained
in the stable regime $m<1$.
}
\label{predv}
\end{figure}

The predictions of the critical time obtained from the three
methods are shown in Figure \ref{predv}.
A prediction for $t_c$ with an uncertainty of a few days is obtained for the
4 benchmarks within 20 days before the catastrophic failure. The reliability
of the prediction is confirmed by the coherence and agreement between
the three methods.
Starting approximately at $t_{\rm max}=45$ days, one can observe that,
using the friction model, all four time series provide a reasonable $t_c$
prediction which however tends to increase and to follow the value of
the ``present time'' $t_{\rm max}$.
This is unfortunately a common feature of fits to power law singularities in
which the last data points close to the ``present'' tends to dominate the rest
of the time series and produce a predicted time of singularity close to the
``present time'' $t_{\rm max}$ \cite{Huangetal00,SornetteJohansen01}.
The $t_c$ value obtained using the fourth-order approximate is always
a little smaller than the $t_c$  estimated from the exact friction model.
The renormalization method is therefore a little better at early times,
but the exact friction model works better at the end.
The $t_c$ value obtained by the linear regression of $1/\dot{\delta}$
is too large for small $t_{max}$, because it is only an asymptotic solution
of the friction model for $t\approx t_c$. However, this method provides
very good estimates of $t_c$ close to $t_c$.

\begin{figure}
\epsfig{file=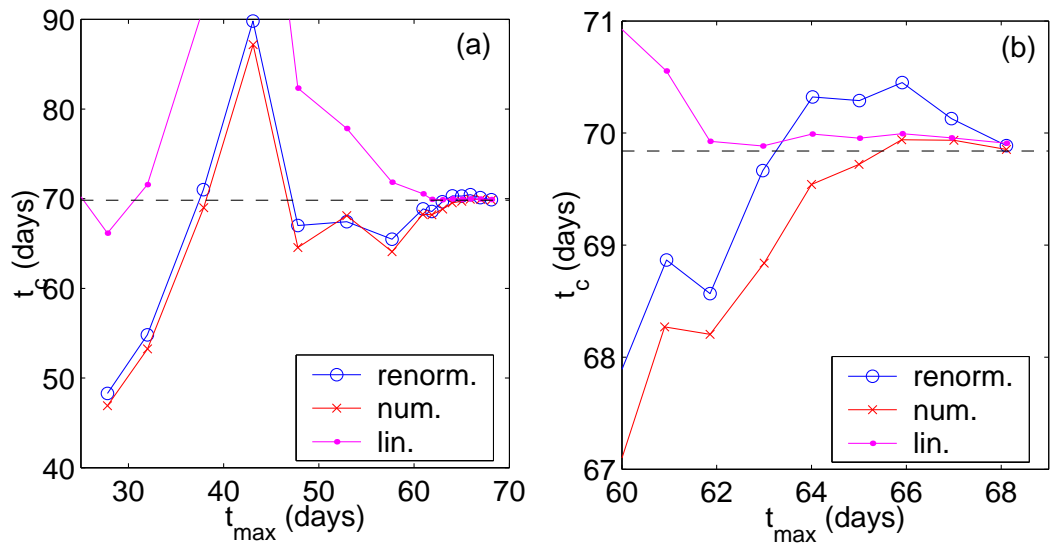,width=14cm} 
\caption{Same as Figure \ref{predv} for a synthetic data set
with the same parameters and noise as those obtained for benchmark 5
of the Vaiont
landslide, using the same three different methods of prediction.
The right panel is a zoom of the left panel close to $t_c$. The horizontal
dashed line indicated the true critical time $t_c=69.8$ of the
catastrophic landslide.
}
\label{comps}
\end{figure}

To test whether the relative value of these three methods result from
a genuine difference in their stability with respect to noise or rather
reflects an inadequacy of the slider-block friction model to fit the data,
we have generated a synthetic velocity time series obtained by using the
slider-block friction equations with the same parameters as found in the
fit to the full data set of benchmark 5 and adding white noise with the
same standard deviation as that of the real data set. We then applied the
three prediction methods to this synthetic data set. In principle and by
construction, we should expect a priori that the prediction based on the
slider-block friction model should always perform best since it is 
the {\it true} model.
This is not what we find, as shown in Figure \ref{comps}. At times far from
$t_c$, i.e. 40 days $<t_{max}<$ 60 days, the friction model
is the best, as expected. However, the
prediction based on the asymptotic linear relation between $1/\dot{\delta}$
and time $t$ is slightly better than the friction model,
starting approximately 9 days before the landslide.

The overall conclusion is that the least sophisticated approach,
that is the linear regression of $1/\dot{\delta}$, seems to perform as well as
or slightly better than the sophisticated renormalization method
or the exact friction model for ``present times'' sufficiently close
to the critical time $t_c$. For times further away from $t_c$, the
renormalization method and the exact friction model are better.
Although the corresponding power-law is only an asymptotic solution
of the friction model for times close to $t_c$,  the linear regression
of $1/\dot{\delta}$ gives significantly better predictions than the exact
model or the renormalization method. However, we must keep in mind that the
use of the linear regression of $1/\dot{\delta}$ as a function of time
contains two hidden and rather strong assumptions: the power law and
the value of its exponent. Without the
slider-block friction model, these assumptions are just guesses
and are a priori unjustified.

\section{Prediction of the aborted 1986-1987 peak acceleration of La 
Clapi\`ere landslide}

\begin{figure}
\epsfig{file=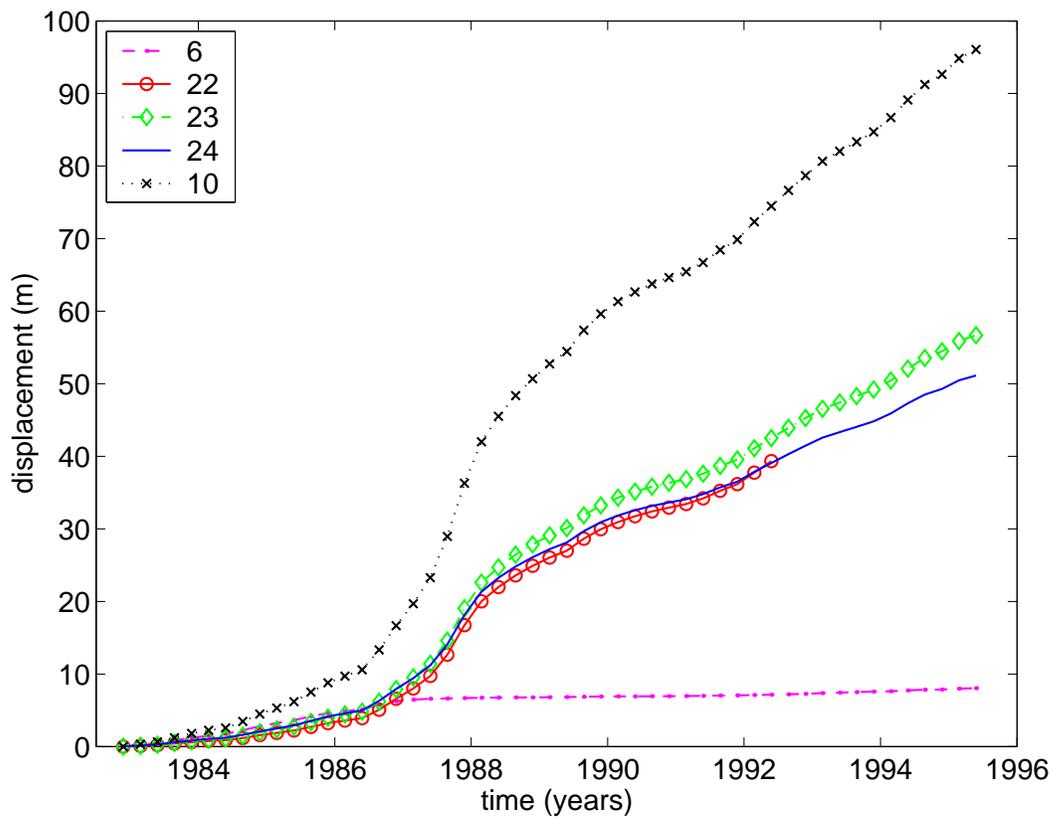,width=14cm}
\caption{Displacement for the 5 benchmarks on La Clapi\`ere used in 
this study.}
\label{4benmarkclap}
\end{figure}

\begin{figure}
\epsfig{file=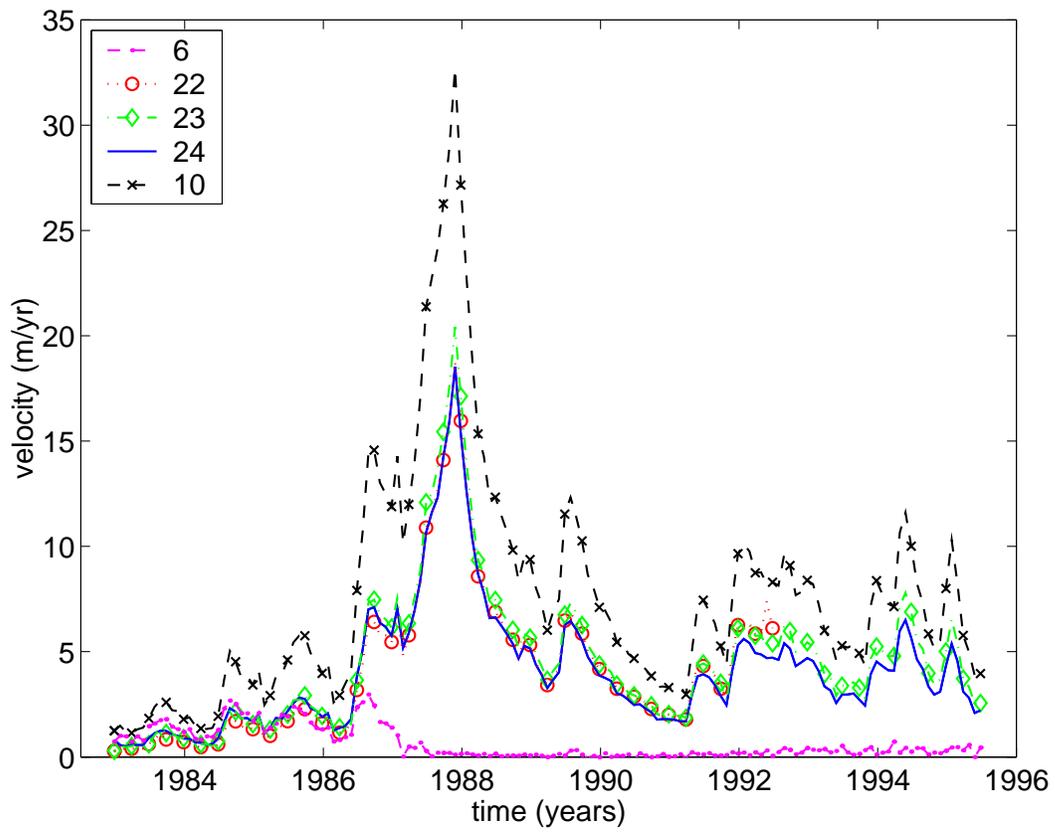,width=14cm}
\caption{Velocity for the same data as shown in Figure
\ref{4benmarkclap}. Annual fluctuations of the velocity is due to
the seasonal variations of the precipitations.}
\label{4benmarkclapv}
\end{figure}

We now report  results on another case which exhibited a
transient acceleration
which did not result in a catastrophic failure but re-stabilized.
This example provides what is maybe an example of the
  $m=B/A<1$ stable slip regime as interpreted within the friction model.
La Clapi\`ere landslide is located at an elevation between 1100 m and 1800 m
on a 3000 m slope high.
The volume of mostly gneiss rocks implied in the landslide
is estimated to be around $50 \times 10^6$ m$^3$. The rock mass started
to be active before the beginning of the 20th century.
The displacement rate measured by aerial photogrametric survey increased from
0.5 m/yrs in the 1950-1960 period to 1.5 m/yrs in the 1975-1982 period
\cite{Follacci88}.
Starting in 1982, the displacements of 43 benchmarks have been monitored
on a monthly basis using distance meters 
\cite{Follacci88,Follacci93,Susella96}.
The displacement data for 5 benchmarks
is shown in Figure \ref{4benmarkclap}.
The velocity is shown in Figure  \ref{4benmarkclapv}.
The rock mass velocities exhibited a dramatic increase between January 1986
and January 1988, that culminated in the 80 mm/day velocity  during
the 1987 summer and to 90 mm/day in October 1987.
The homogeneity of benchmark trajectories and the synchronous 
acceleration phase
for most benchmark, attest of a global deep seated behavior of this landslide
\cite{Follacci88}.
However, a partitioning of deformation occurred, as reflected by the
difference in absolute values of benchmark displacements (Figure 
\ref{4benmarkclap}).
The upper part of the landslide moved slightly faster than the
lower part and the NW block.
The observed decrease in displacement rate since 1988 attest of a change
in landsliding regime at the end of 1987 (Figure \ref{4benmarkclap}).

\subsection{Correlations between the landslide velocity and the river flow}

\begin{figure}
\epsfig{file=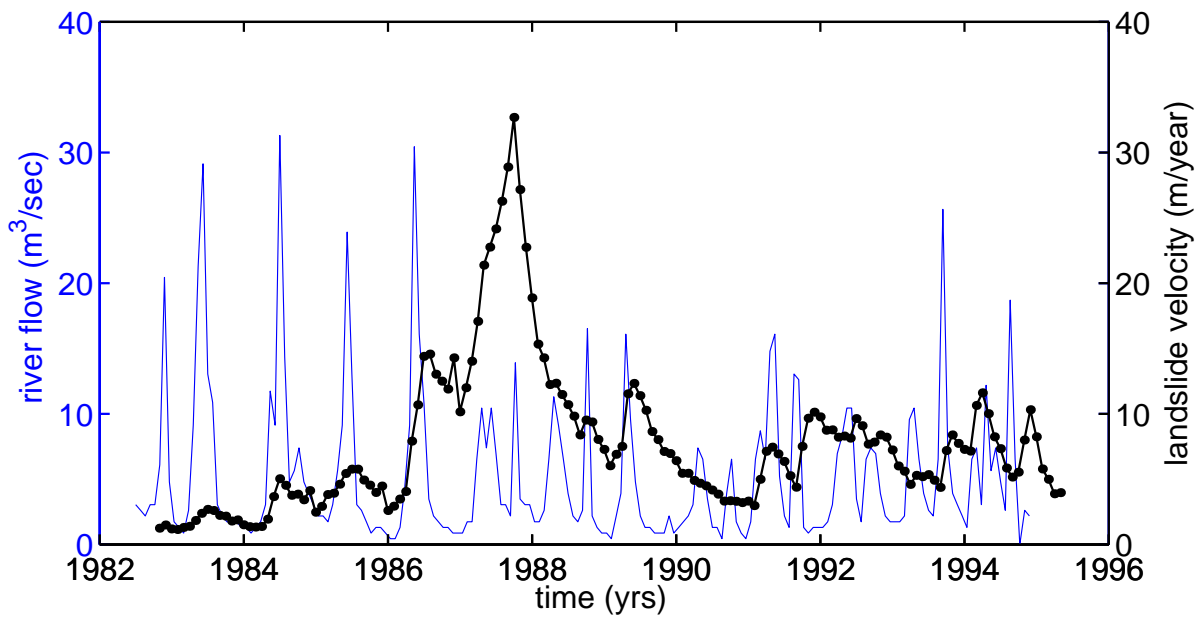,width=16cm}
\caption{Velocity pattern for benchmark 10 of  La Clapi\`ere 
landslide (solid line and dots)
and flow rates (thin solid line) of
the Tin\'ee river on the 1982-1995 period. Because the Tin\'ee river 
runs at the
basis of the La Clapi\`ere landslide, the river flow rate reflects the
water flow within the landslide \cite{Follacci93,Susella96}.
  The flow rates are measured at St Etienne village, 2 km upstream the 
landslide site.
There is no stream network on the landslide site. The Tin\'ee flow 
drains a 170 km$^2$ basin.
This tiny basin is homogeneous both in terms  of slopes and elevation 
(in the 1000-3000 m range).
Accordingly the seasonal fluctuations of the river flow is admitted 
to reflect
the evolution of the amount of water that is available within the landslide
slope due to rainfalls and snow melting. Data from {\it CETE}, [1996].
}
\label{debitpluie}
\end{figure}

The velocity displays large fluctuations correlated with fluctuations of
the river flow in the valley as shown in Figure \ref {debitpluie}.
There is a seasonal increase of the slope velocity which reaches a maximum
$V_{\rm max}$ of the order of or less than $30$ mm/days.
The slope velocity increases in the spring due to snow melting and over
a few days after heavy precipitations concentrated in the fall of each year
\cite{Follacci88,Susella96}.
During the 1986-1988 period, the snow melt and rainfalls were not 
anomalously high
but the maximum value of the velocity, $V_{\rm max} = 90$ mm/day, was 
much larger
that the velocities reached during the 1982-1985 period for comparable
rainfalls and river flows \cite{Follacci88,Follacci93}.
This strongly suggests that the hydrological conditions are not the 
sole control
parameters explaining both the strong 1986-1987 accelerating and the 
equally strong
slowdown in 1988-1990.
During the interval 1988-1990, the monthly recorded velocities slowed down
to a level slightly higher than the pre-1986 values. Since 1988, the seasonal
variations of the average velocity never recovered the level established during
the 1982-1985 period \cite{Follacci93,David00}.
{\it Rat} \cite{Rat88} derives a relationship between the river flow and the
landslide velocity by adjusting an hydrological model to the velocity data
in the period 1982 to 1986. This model tuned to this time period
does not reproduce the acceleration of the velocity after 1986.

In order to study quantitatively the effect of the precipitations on 
the landslide velocity,
we need to remove the long-term fluctuations of the velocity that may not be
correlated to changes in the precipitations.
Before applying a spectral analysis of the velocity data, we
use simple functions to fit the displacement data. We then subtract
this long-term trend to obtain stationary residuals that can be used
to perform a spectral analysis of the fluctuations of the velocity.
We divide the data of benchmark 10 of La Clapi\`ere
into three different intervals: $[1982.917, 1987.833]$,
$[1987.833, 1991.25]$ and $[1991.25, 1995.5]$.
The initial values of the time and of the displacement are fixed to 0 
at the beginning
of each time period.
In the first interval,
the velocity rises (with
fluctuations); in the second interval, the velocity decreases (with
fluctuations); in the third interval, the velocity fluctuates around a
constant. We used non-linear Least-Square fits with different
fitting functions separately within each interval.
The results of the fits are the following.
\begin{enumerate}
\item In the first interval $[1982.917, 1987.833]$, we fit the displacement
by $d(t) = a (|1-t/t_0|^{-b} - 1)$ with  $a=8.96, b=1.01$ and $t_0=6.26$ years.

\item For the second interval $[1987.833, 1991.25]$, we use the same functional
form with $a=10.42,  b=0.4106$ and $t_0= -0.1081$. The negative value of $t_0$
implies a decay of the displacement.

\item  For the third interval $[1991.25, 1995.5]$, we use a  fit by
$d(t) =a t^b$ which has only two parameters $a=7.4687$ and $b=0.989$.
\end{enumerate}
The goodness of fit is very good in all three regimes: the standard
deviations of the residuals being of the order of $0.4$ while the
magnitude of the displacement is about $30$, this yields a signal-over-noise
ratio of $75$, which is very good.

\begin{figure}
\epsfig{file=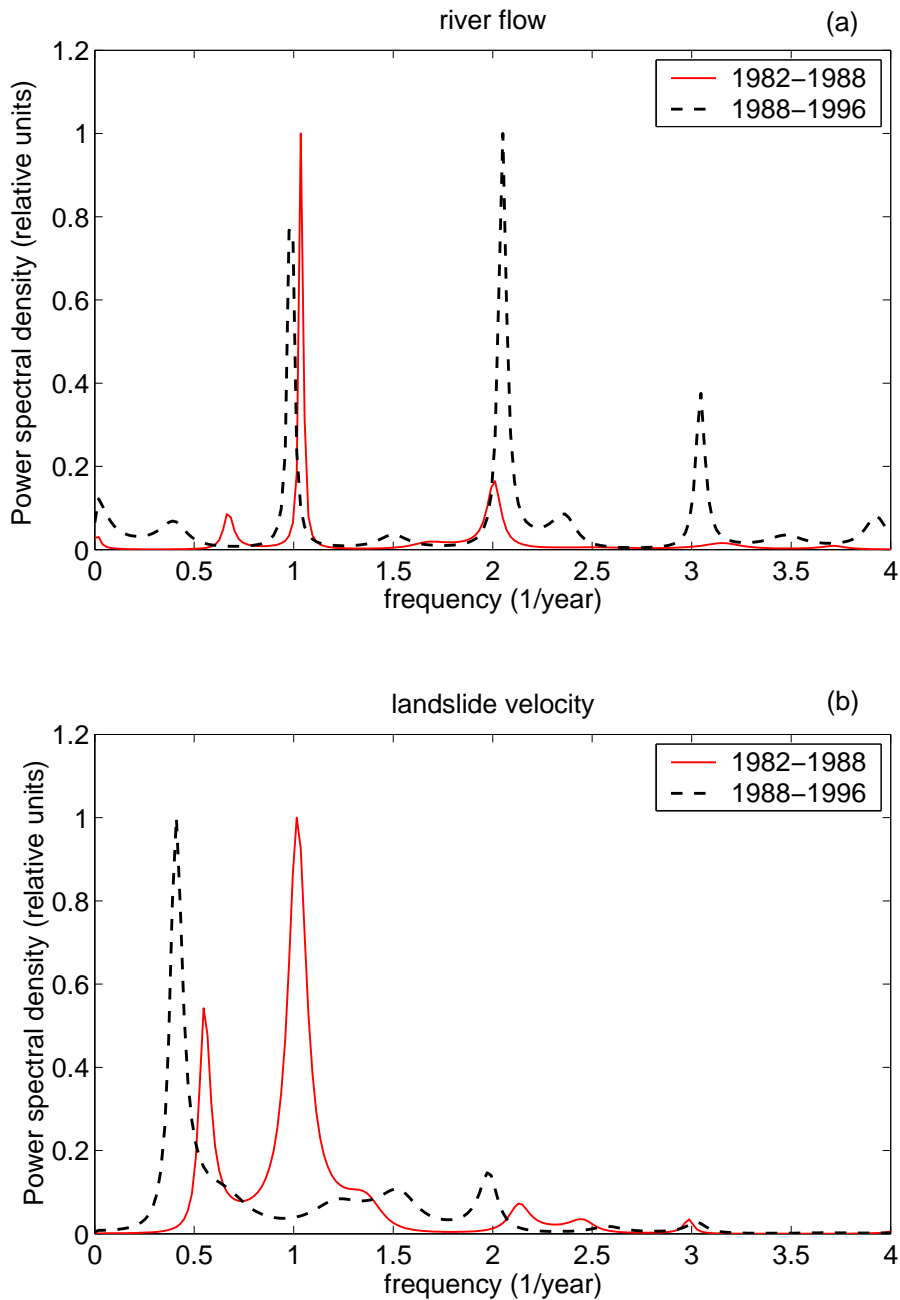,width=12cm}
\caption{Top panel: Burg's power spectrum of the
flow rates of the Tin\'ee river on the 1982-1988 and on the
1988-1996 periods which are aggregated from the periods shown in 
figure \ref{debitpluie}.
Bottom panel: Burg's power spectrum of
the detrended velocity residuals for the same two periods.
}
\label{FIG2Pisa}
\end{figure}

Figure \ref{FIG2Pisa} compares the Burg's power spectrum of the
flow rates of the Tin\'ee river and of the detrended velocity residuals.
The Burg spectrum is a smoothed FFT (fast-Fourier transform)
obtained by approximating the true spectrum by that
of an autoregressive process of a finite order.
The top panel of figure \ref{FIG2Pisa} exhibits the Burg's power 
spectrum of the
flow rates of the Tin\'ee river on the 1982-1988 and on the 1988-1996 
periods, which
are proxies of the cycle of precipitations and snow melting.
The bottom panel of figure \ref{FIG2Pisa} shows the Burg's power spectrum
of the detrended velocity residuals for these two periods.

In the first time interval 1982-1988, a strong peak
at the period of $1$ year appears both for the
velocity residuals and for the river flow. This correspondence is 
confirmed by the strong
cross-correlation between the river flow and the landslide velocity, 
which is also
directly apparent visually in Figure \ref{debitpluie}. We now use the language
of system theory and consider the river flow
as an input (or a forcing) and the landslide velocity as an output of the
system. These observations of a common spectral peak and of a strong 
cross-correlation
are then compatible with a view of the system as being linear or only 
weakly non-linear.

In contrast, the (linear)
correlation between the river flow input and the landslide velocity
output disappears in the second time interval 1988-1996,
as can been seen from the absence of a
spectral peak at the period of $1$ years and a very weak
peak at the period $6$ months ($f=2$ year$^{-1}$)
in the (output) landslide velocity spectrum compared with the two 
strong peaks at
the same periods of
$1$ years and $6$ months observed in the (input) river flow spectrum.
This breakdown of linear correlation seems to be associated with the birth of
a strong peak close to the sub-harmonic period of $2$ years ($f=0.5$ 
year$^{-1}$),
which is absent in the river flow rate.
This suggests the following interpretation. Frequency doubling or
more generally frequency multiplications are the results of simple 
nonlinearities.
Indeed, higher frequency overtones in river runoff is very common feature of
hydrological regime \cite{Pisa01}. In contrast,
the creation of sub-harmonics requires bifurcations or 
period-doubling, for instance
involving nonlinear processes with time delays.
It thus seems that the input of rain
and snow melting is transformed by the system during the second
time interval via the process of such delayed period-doubling nonlinearities.
It is intriguing
that the change of sliding regime to a reduction of velocity in the 
second time interval
seems here to be associated with such a sub-harmonic non-linearity, 
which could be
the result of a change of topology of the block structures (through 
fragmentation)
and of the solicitation of novel fresh surfaces of sliding.

It would also be interesting to add a
periodic forcing to our models to better capture the time-dependence of
the velocity and study its possible nonlinear consequences.
This is left for a future work, together with a complete
description of the three time intervals by the slider-block friction model.

\subsection{Prediction of La Clapi\`ere change of sliding regime in 1983-1988}

\begin{figure}
\epsfig{file=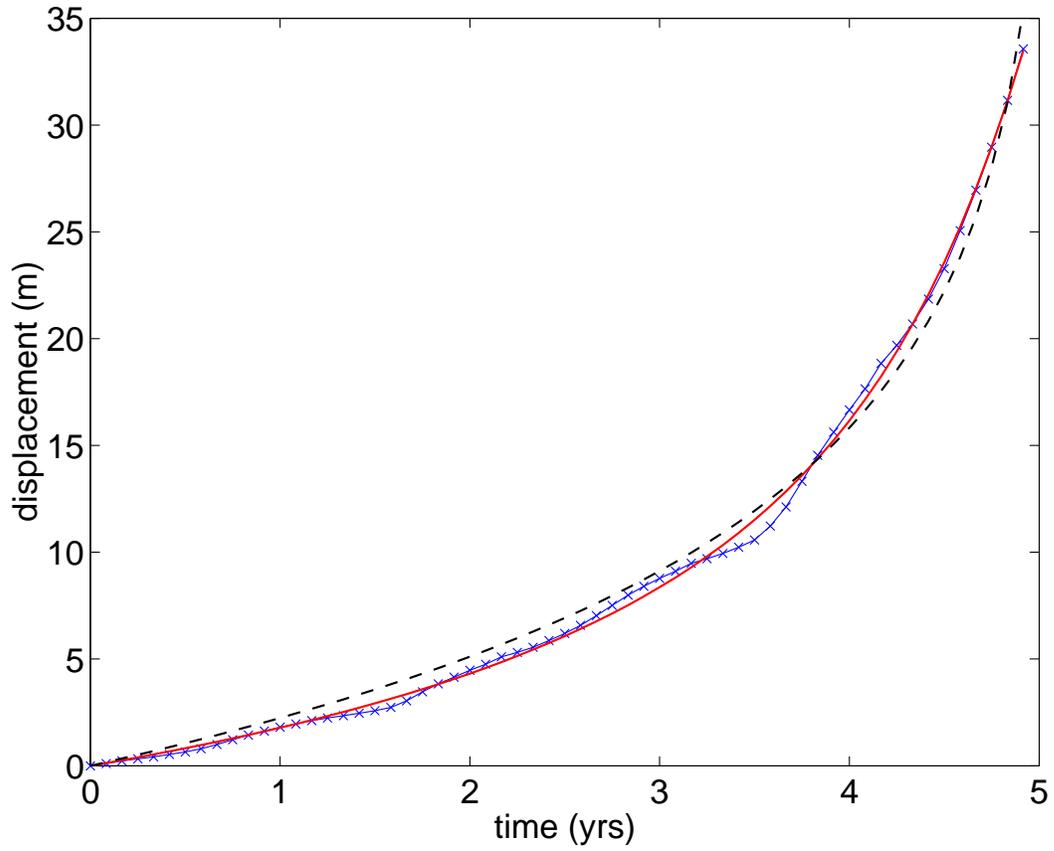,width=14cm}
\caption{ Displacement for benchmark 10 of la Clapi\`ere landslide (crosses)
and fit using the friction model.
The best fit gives $m=0.98$ (black line). The gray line shows the best fit
obtained when imposing $m=1.5$ for comparison.}
\label{depclap}
\end{figure}

\begin{figure}
\epsfig{file=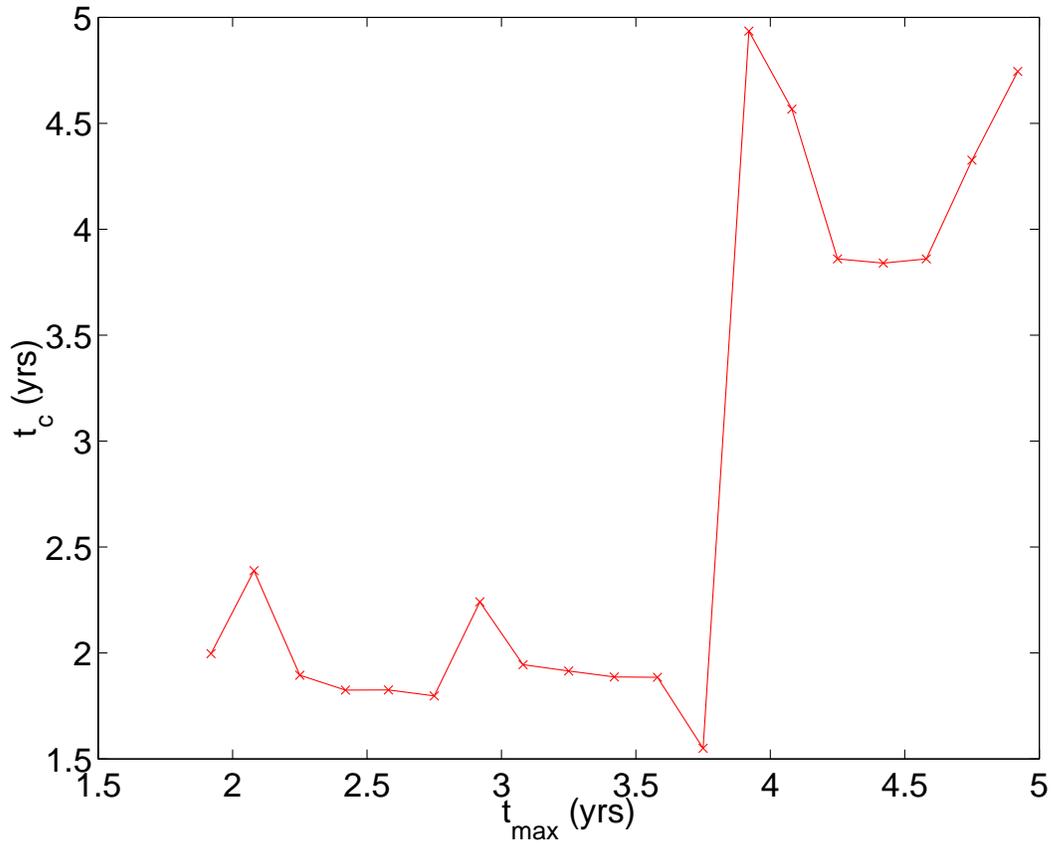,width=14cm}
\caption{Predicted value of the time $t_c$ of the inflection point of
the velocity for La Clapi\`ere  landslide, using a fit of the displacement
data with the friction model.
All points correspond  to the stable regime $m<1$.
In this regime there is no finite-time singularity of the velocity
but a transition from an accelerating sliding to a stable sliding for
times larger than the inflection point $t_c$.
This parameter is poorly constrained by the fit and increases
with the time of the last point $t_{max}$ used in the fit.}
\label{tctmaxclap}
\end{figure}

Our previous analysis of the calibration of the frictional model to the
displacement of La Clapi\`ere data finds that
$m=B/A$ is very close to but smaller than one,
while the value of $x_i$ is significantly larger
than 1. The corresponding fit of the displacement data with the 
slider-block model
is shown in figure \ref{depclap}.
This argues for La Clapi\`ere landslide to be in the stable regime
\cite{comphelmjgr}. However, the transition time (defined by the inflection
point of the displacement) is found to increase with $t_{\rm max}$
as shown in figure \ref{tctmaxclap}. This may argue for a change of regime
from an acceleration regime to a restabilization before the time $t=1988$
of the velocity peak (corresponding to the inflection point). The
parameters $S$ and $x_i$ found in this analysis are also poorly constrained.
Similar results are obtained for different benchmarks.

\begin{figure}
\epsfig{file=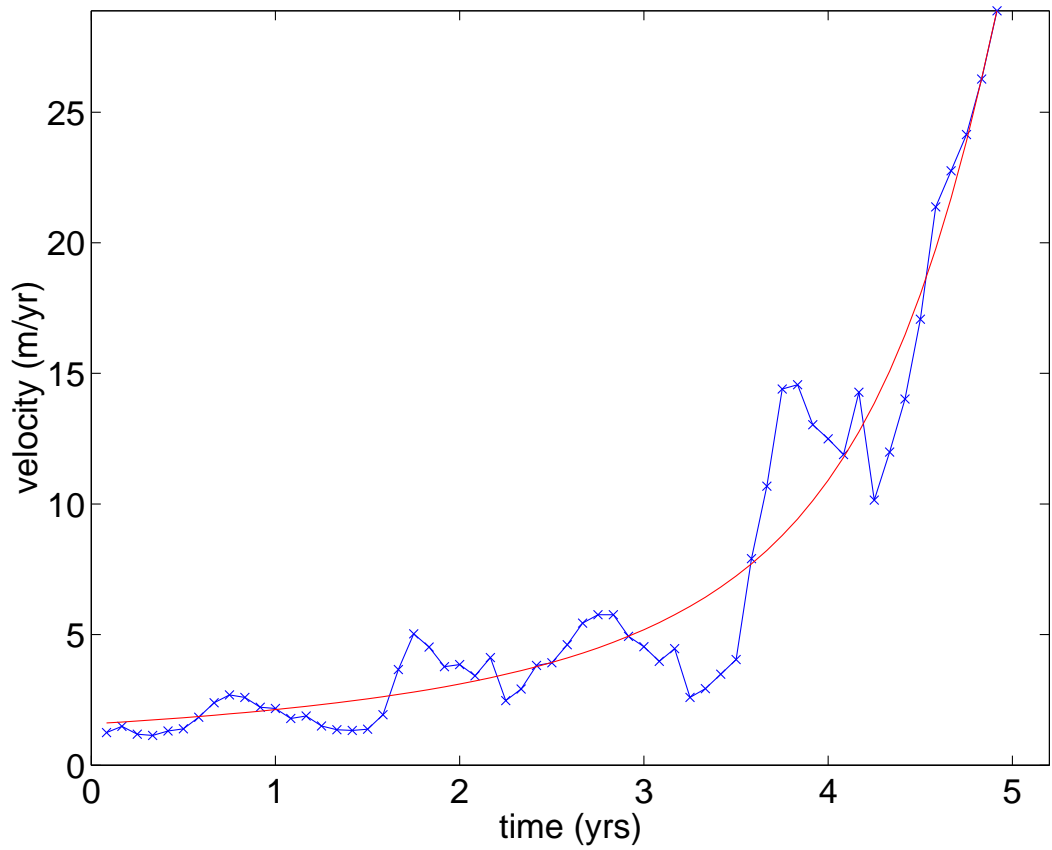,width=14cm}
\caption{Velocity for benchmark 10 of la Clapi\`ere landslide (crosses)
and fit of the velocity data with the friction model.
The best fit gives $m=0.99$ (black line).}
\label{vclap}
\end{figure}

\begin{figure}
\epsfig{file=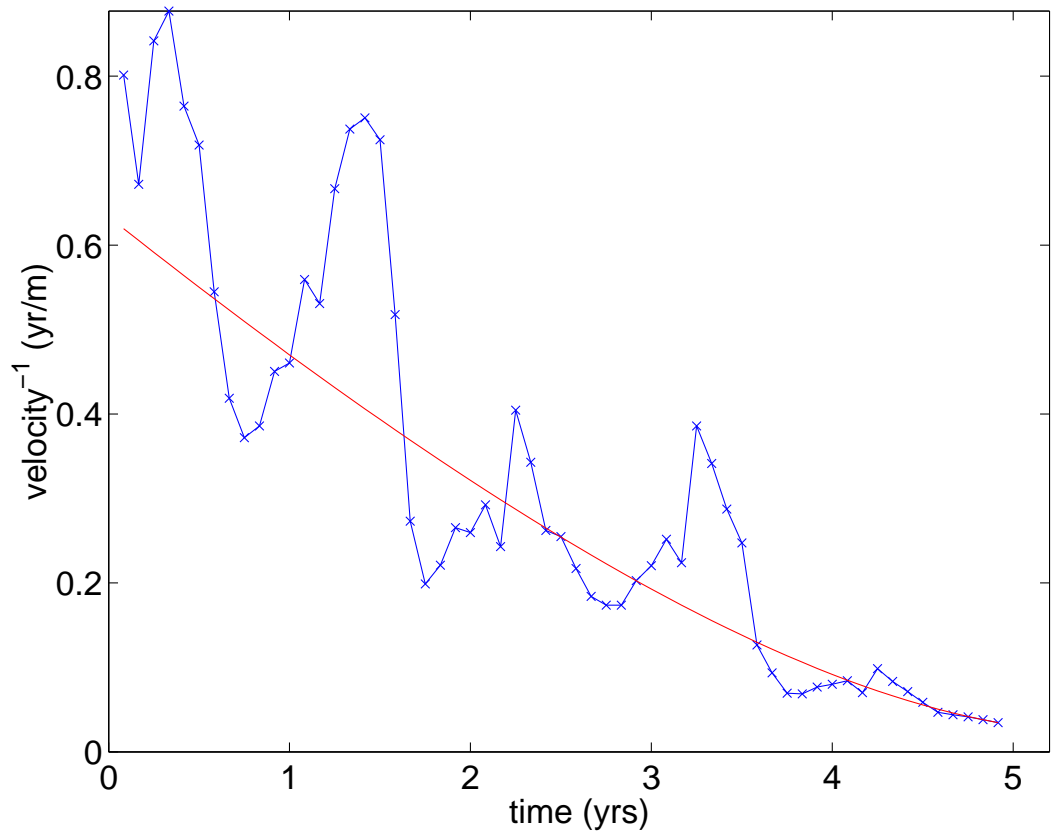,width=14cm}
\caption{Same as Figure \ref{vclap} showing the inverse of the velocity.}
\label{vm1clap}
\end{figure}

The analysis of the velocity data seems to reinforce
somewhat the idea of a change of regime from an unstable to a stable 
phase, as shown in
figures \ref{vclap} and \ref{vm1clap}:
the early acceleration was in the unstable
regime $m>1$ but did not reach the instability due to a change of morphology,
block partition and the creation of new active surfaces of sliding.
This interpretation is suggested in particular by the plot of the 
inverse of the velocity shown
in Figure \ref{vm1clap}, which is close to linear at early times.
Over the route toward the finite-time singularity, the landslide
perhaps did not
succeed in accommodating the velocity increase and degenerated
by changing geometry and loading conditions (block partitioning).
In other words, the solution shown in Figure \ref{depclap}
with $m<1$ may rather describe a transient from an unstable state to 
a stable regime.
In particular, we cannot exclude the possibility that the surfaces 
have all along been
characterized by the regime $B>A$ and then a change of geometry
and surfaces of sliding may have reset the reduced state variable $x$ given
by expression (\ref{mgmbm}).
Another possibility is that the friction parameter $m$ has changed 
from $m>1$ to $m<1$,
leading to a stable deceleration of the displacement after 1988.
It is not unreasonable to conjecture that  the internal stresses associated
with and created by the accelerating phase
may have led to its fragmentation into several sub-entities, creating
fresh surfaces and resetting the state variable or the $m$-value characterizing
the surfaces of contact. This is in qualitative agreement with field 
observations of new faulting patterns since 1987, which signal a change 
in the geometry of the landslide involving the regression of the main 
scarp and locked sub-entities \cite{Follacci93,GuVe02}.
These observations provide evidence for a change in both the head
driven force (mass push from the top) and the activated basal surfaces.
These morphological changes suggest that the 1987-1988 period has been a
transition period for the evolution of La Clapi\`ere sliding system over
the last 50 years.
In the block-slider model, this amounts to modifying the variables $S$
and $\theta_i$ and thus to reset $x$. In this interpretation, the change of
regime observed for La Clapi\`ere could then be due to a change from
  $x_i<1$ (unstable acceleration) to $x_i>1$ (stable deceleration).
This change from $x_i<1$ to $x_i>1$
may be interpreted as either an increase of applied shear stress, a
decrease of normal stress, or an increase of the surface of contacts
between the sliding surfaces.
Thus, within the slider-block model, one can characterize
the post 1988 landslide evolution in terms of new sliding surfaces
being mobilized which are more stable that the previous ones due to
more numerous and/or efficient contacts.

Appendix B explores what would have been the predicted critical
time $t_c$ estimated in real time prior to the velocity peak, according
to this scenario of an unstable acceleration towards a finite-time
singularity. We have seen that, while the slider-block model as well as
the power law formula (\ref{jmgjkla}) provide excellent fits to the data, they
do not lead to very stable predictions of the critical time $t_c$ on the Vaiont
data as well as on synthetic tests generated in the unstable regime $m>1$.
It may thus be valuable to
test the approach of Gluzman et al. \cite{Gluzmanetal01} in terms of 
a version of the functional
renormalization approach already discussed in relation with the 
Vaiont landslide.
It is our hope that this approach could provide in a more robust 
determination of $t_c$.

Figure \ref{renorm} compares the prediction of a fit using a 
polynomial of order two in time
to the inverse of the velocity (panel (a)) with the prediction
of the renormalization approach (panel (b)). In each panel, two 
curves are presented
corresponding to two different starting points of the data taken into 
account in the
predictions: the points to the left correspond to the
first date taken into account in the predictions; therefore, the predictions corresponding
to the crosses $\times$ use approximately two years fewer 
data than the predictions shown
with the open circles. This allows us to compare the effect of missing data
or alternatively the effect of a non-critical behavior at the
beginning of the time series.
The abscissa $t_{\rm max}$ is the running ``present time'', that is, the last
time of the data taken into account to issue a prediction.
The prediction with the polynomial shown in panel (a) of Figure \ref{renorm}
can be seen as an improvement in methodology over the Voight formula 
(\ref{jmgjkla})
which corresponds to a linear fit of the inverse velocity with time 
for $\alpha =2$.
Comparing panels (a) and (b), the renormalization method seems to 
present a smaller
dispersion and better convergence: in particular, about half-a-year 
prior to the time of the
maximum realized velocity indicated by the horizontal dashed line, 
the prediction of this
date by the renormalization method using the longer time series
becomes very precise. Thus, a critical time close to the time of the 
velocity peak
would have been predicted starting approximately half-a-year year from it.
It is then not unreasonable to consider the velocity peak as a proxy for the
critical time that the system would have exhibited in absence
of a change of regime, since on its approach the largest
internal stresses may develop and may fragment the block and modify the
morphology of the landslide, thus resetting the geometry and some of the
parameters of the model. In this scenario, we would thus expect that the time
of the peak velocity should be not far from what would have been the critical
time of catastrophic failure of the landslide.

We should however point out
that the functional renormalization method
used in this Appendix B does not work for the Vaiont landslide
because of a technical instability whose fundamental origin is not 
understood by these authors.
Technically, the numerical
instability comes from the absence of alternating signs in the polynomial
expansion at early times. This technical problem thus casts some shadow on the
usefulness of the approach described here which is unable to tackle
the regime which is undoubtedly unstable. This limitation suggests again the
importance of working with several alternative and competing models,
as further discussed in the following concluding section.

\section{Discussion and conclusion}

We have extended the quantitative analysis of our companion
paper \cite{comphelmjgr} on the displacement history for
two landslides, Vaiont and La Clapi\`ere, to explore their potential
predictability. Using a variety of techniques,
we have tried to go beyond the time-independent hazard analysis provided
by the standard stability analysis to include time dependent predictions.
While our present inversion methods provide a single estimate of the
critical time $t_c$ of the collapse for each inversion, a better formulation
should be to translate these results in terms of a probability of failure, as
for instance done by Vere-Jones et al. \cite{Vere01}.

Using the innovative concept of applying to landslides
the state and velocity dependent friction law established in the laboratory
and used to model earthquake friction, our inversion of this simple
slider-block friction model shows that the observed movements can be well
reproduced and suggest the Vaiont landslide
(respectively La Clapi\`ere landslide) as belonging to the velocity
weakening unstable (respectively strengthening stable) regime.

For the Vaiont landslide, the friction model provides good predictions
of the time-to-failure up to 20 days before the collapse.
A pure phenomenological model suggested by Voight \cite{Voight88}
postulating a power law finite-time
singularity $\dot{\delta} \sim 1/(t_c-t)$ with unit exponent obtains
similar results up to 10 days before the collapse.
Our approach can be seen as providing a physically-based
derivation of this phenomenological model as well as a generalization
to capture three other possible regimes.

For la Clapi\`ere landslide, the inversion of the displacement data
for the accelerating phase 1983-1888 up to the maximum of the velocity
gives $m<1$, corresponding to the stable regime.
The deceleration observed after 1988 implies that, not only
is la Clapi\`ere landslide in the stable regime but in addition,
some parameters of
the friction law have changed, resulting in a change of sliding regime
from a stable regime to another one characterized by a smaller velocity,
as if some stabilizing process was occurring.
Possible candidates for a change in landsliding regime include the average
dip slope angle, the partitioning of blocks, new sliding surfaces and changes
in interface properties.
However, another possible interpretation  is that this landslide
was initially in the unstable regime, but did not reach the instability
due to a change of geometry and of sliding surfaces.
The best fit obtained with $m<1$ for the accelerating phase 1983-1988
would then describe a transient regime between the unstable regime and
the stable regime, due to a progressive change in the model parameters.
This second scenario seems less parsimonious but cannot be completely excluded.

The present work has offered the novel
conceptual framework and language of the slider-block model,
which can be used to classify the relative merits and performance of other
models. For an assessment in real time of the upcoming risks of a catastrophic
failure, one should then consider both scenarios (stable versus unstable
which are encoded respectively by the range of parameters
$m<1$ and $m>1$ in the slider-block model) and test the data using the
available associated theoretical models, some of which have been presented
in this paper. Such an approach in terms of multiple scenarios
\cite{Smith99,YukalovGluzman99,Ziehmann00}
can help assess societal risks. A systematic exploration of such approaches
will extend the preliminary investigation and results offered here.

\vskip 0.3cm
{\bf Acknowledgments}:
We thanks C. Scavia and Y. Guglielmi for key supports to capture 
archive data for
  Vaiont and La Clapi\`ere  landslide respectively.
  We are very grateful to N. Beeler, J. Dieterich, Y. Guglielmi,  D. Keefer,
J.P. Follacci, J.M. Vengeon  for useful suggestions and discussions.
AH and JRG were supported by INSU french grants, Gravitationnal 
Instability ACI.
SG and DS acknowledge support from the James S. Mc Donnell Foundation 21st
century scientist award/studying complex systems.

\appendix

\section{Appendix A: Functional renormalization group formulas for the
friction law  (\ref{velocity}) and (\ref{ode})}

Consider an expansion as in (\ref{6}) of an observable $x(t)$ in powers of a
variable $u$ given by $x_k(u)=\sum_{n=0}^ka_n~u^n$. The method of algebraic
self-similar renormalization constructs so-called ``approximants'', which
are reconstructed functions that best satisfy the imposed asymptotic
constraints while obeying criteria of functional self-similarity and
of maximum stability in the space of functions
\cite{YukalovGluzman97,GluzmanYukalov97,GluzmanYukalov98}.
These approximants are given by the following general recurrence formula
for the approximate $x_k^{*}(u)$ of order $k$ as a function of the
expansion $x_{k-1}(u)$ up to order $k-1$ :
\be
\label{12}
x_k^{*}(u) = \left[ x_{k-1}^{-k/s}(u)-\frac{k~a_k}s~u^k\right]^{-s/k}~.
\ee
The crossover index $s$ is determined by the condition that the
leading terms of the expansion of $x_k^{*}(t)$ as $t\rightarrow 0$
must agree with the expansion of $x_k(u)$.

For the friction model (\ref{velocity}) and (\ref{ode}), the 
coefficients $a_k$ in
(\ref{6}) and (\ref{12})
are determined by the friction parameters and the initial conditions
\ba
\label{7}
a_0 &=& y_0, \\
a_1 &=&\theta_0^{-1}-S \ y_0^{1-m},\\
a_2 &=&\frac{1}{2} S\ (m-1)\ a_1 y_0^{-m}~,\\
a_3 &=& \frac 16\alpha (m-1)\left[ -m\ a_1^2\ y_0^{-m-1}+2a_2\
	x_0^{-m}\right] ~,\\
a_4 &=&\frac{1}{24} S (m-1)~ [ \left( 1+m\right) \ m\ a_1^3\ y_0^{-m-2}
\nonumber \\
   &~& ~ -6m\ a_2\ a_1 y_0^{-m-1}+6a_3\ y_0^{-m}] ~,
\ea
where $y_0=\theta(t=0)/\theta_0$.

\subsection{Case $m<1$}

As we see from (\ref{10}), the natural expansion variable is
$u=\exp \left( -\frac t{t^{*}}\right)$.

The first-order and simplest approximate is
\begin{equation}
\label{12bis}x_1^{*}(u)=x^{*}\left( 1+cu\right) ^{-s}=x^{*}\left( 1+c\exp
\left( -\frac t{t^{*}}\right) \right) ^{-s}~,
\end{equation}
with $x^{*} = 1/T$ where $T$ is given by (\ref{mgmlla}).
The crossover amplitude $c$ and the crossover index $s$ are determined
by the condition that the expansion of $x_1^{*}(t)$ as $t\rightarrow 0$
must agree with the first two terms of expression (\ref{6}), leading
to the following
system of equations,
\begin{equation}
\label{14}x^{*}\left( 1+c\right) ^{-s}=x_0~,
\end{equation}
\begin{equation}
\label{15}x_0s{\ }c\ \frac 1{t^{*}(1+c)}=a_{1} ~.
\end{equation}
The crossover index $s$ is then given by
\be
s=-\frac{\ln \left( x_0/x^{*}\right) }{\ln \left( 1+c\right) }~,
\ee
while the crossover amplitude $c$ satisfies the following equation:
\be
\frac{\ln \left( x_0/x^{*}\right) }{\ln \left( 1+c\right) }\ \frac
c{(1+c)}=- \frac{a_1t^{*}}{x_0}~.
\ee

The second-order approximate is given by
\ba
\label{16}x_2^{*}(u) &=& x^{*}\left[ \left( 1+c_2u\right) ^{-s_2}+
c_1u^2\right] ^{-s_1} \nonumber \\
&=& x^{*}\left[ \left( 1+c_2\exp \left( -\frac t{t^{*}}\right) \right)
^{-s_2}+c_1\exp \left( -\frac{2t}{t^{*}}\right) \right] ^{-s_1} ~.
\ea
The crossover amplitudes $c_1,c_2$ and crossover index $s_1$
and $s_2$ are obtained from the condition that the expansion of $x_2^{*}(t)$ as
$t\rightarrow 0$ must recover the first four terms of expression (\ref{6}).
The corresponding expressions are rather long and will not be presented here
explicitly. Interestingly, for $m=0$, the second-order approximate
recovers the exact solution.

\subsection{Case $m\geq 1$}

In this case, the natural variable in the expansion is $u=t$. Our goal
is to obtain the critical time $t_c$ as a function of $m$. Using the crossover
technique \cite{GluzmanYukalov98} for the two asymptotic expressions
(\ref{6}) at short time and (\ref{mmsls}) close to $t_c$,
we obtain a sequence of approximants
$x_1^{*}(t),\ x_2^{*}(t),x_3^{*}(t)$ and $x_4^{*}(t)$ associated with
a sequence
of improving approximations for the critical time, $t_{c1}(m),\
t_{c2}(m),\ t_{c3}(m)$
and $t_{c4}(m)$. All approximants agree term-by-term with the corresponding
short time expansion and lead to the critical behavior (\ref{mmsls})
as $t$ goes to
the corresponding critical time. The first-order approximate is
\begin{equation}
\label{17}x_1^{*}(t)=x_0\left( 1+\frac{a_1}{x_0}m\ t\right) ^{1/m},\
{\rm with}\ t_{c1}=-\frac{x_0}{m\ a_1}~.
\end{equation}
Interestingly, $x_1^{*}(t)$ coincides with the exact solution in the limit
$m\rightarrow \infty$, which takes the form
$x= x^* \left( (x_0/x^*)^m-(t/t^*)\right)^{\frac 1m}$.

In the next order, we obtain the second-order approximate
\begin{equation}
\label{18}x_2^{*}(t)=x_0\left[ \left( 1+\frac{a_1}{x_0}\ t\right) 
^m+\ \frac{
m\ a_2}{x_0}\ t^2\right] ^{1/m}~,
\end{equation}
and $t_{c2}$ is solution of the following equation
\begin{equation}
\label{19}\left( 1+\frac{a_1}{x_0}\ t_{c2}\right) ^m+\ \frac{m\ a_2}{x_0}\
t_{c2}^2=0~.
\end{equation}
The third order approximate reads
\begin{equation}
\label{20}x_3^{*}(t)=x_0\left[ \left( 1+\frac{a_1}{x_0}t+\frac{a_2}{x_0}%
t^2\right) ^m+\ \frac{m\ a_3}{x_0}t^3\right] ^{1/m},
\end{equation}
and $t_{c3}$ satisfies the following equation
\begin{equation}
\label{21}\left( 1+\frac{a_1}{x_0}\ t_{c3}+\frac{a_2}{x_0}t_{c3}^2\right)
^m+\ \frac{m\ a_3}{x_0}\ t_{c3}^3=0~.
\end{equation}
The fourth-order approximate is given by
\begin{equation}
\label{22}x_4^{*}(t)=x_0\left[ \left( 1+\frac{a_1}{x_0}t+\frac{a_2}{x_0}t^2+
\frac{a_3}{x_0}t^3\right) ^m+\ \frac{m\ a_4}{x_0}t^4\right]^{1/m}~,
\end{equation}
with $t_{c4}$ solution of the equation
\begin{equation}
\label{23}\left( 1+\frac{a_1}{x_0}t_{c4}+\frac{a_2}{x_0}t_{c4}^2+\frac{a_3}{%
x_0}t_{c4}^3\right) ^m+\ \ \frac{m\ a_4}{x_0}t_{c4}^4=0~.
\end{equation}

Note that for $m=1$, all approximants are identical and equal to the
exact solution.

\section{Appendix B: Functional renormalization of polynomials expansions
for the prediction of $t_c$ as a function of the ``present time'' $t_{\rm max}$
for La Clapi\`ere landslide}

This appendix present tests of the prediction of the time at which the velocity
peaked, following the hypothesis discussed in the main text that
the ensuing deceleration resulted from a change from $x_i<1$ to $x_i>1$
in the velocity weakening regime $B>A$. According to this interpretation,
the first accelerating phase should be described by an increasing velocity
$\propto 1/t_c-t)$. The critical time $t_c$ can be approximated by the
time of the peak of the velocity, in other words, $t_c$ is close to
the inflection point of the
displacement as a function of time.

Rather than using the version of the functional renormalization method
described for the Vaiont landslide based on the slider-block equations
of motion, we use here a simpler version
that has been tested earlier in another rupture problem
\cite{Gluzmanetal01}. This choice is
governed by the fact that we can not rely entirely on the friction model
with fixed parameters since we know that a change of regime occurred. We thus
follow a more general approach which is not dependent upon a specialized
specification of the equations of motion.
The previous investigation on a model system \cite{Gluzmanetal01}
developed theoretical formulas for
the prediction of the singular time of systems which are a priori known to
exhibit a critical behavior, based solely on the knowledge of the early time
evolution of an observable. From the parameterization of such early time
evolution in terms of a low-order polynomial of the time variable, the
functional renormalization approach introduced by Yukalov and Gluzman
\cite{YukalovGluzman97}
allows one to transform this polynomial into a function which is
asymptotically a power law. The value of the critical time $t_c$,
{\it conditioned} on the assumption that $t_c$ exists, can then be determined
from the knowledge of the coefficients of the polynomials. {\it Gluzman et al.}
\cite{Gluzmanetal01} have tested with success this prediction scheme 
on a specific
example and showed that this approach gives more precise and reliable
predictions than through the use of the asymptotic power law model, but is
probably not better than the true model when the later is known.

The input of the method is the inverse of La Clapi\`ere block velocity
~$\dot{\delta}$ as a function of
time up to the ``present time'' $t_{\rm max}$. One starts with a simple
polynomial fit of $1/\dot{\delta}$ as a function of time from some
starting time up to $t_{\rm max}$. One then applies the functional
renormalization method explained in \cite{Gluzmanetal01} to this
polynomial expansion. We restrict our analysis to expansions of up to
second-order in time:
\be
1/\dot{\delta} = 1 + b_1 t + b_2 t^2~,
\label{mgmt2thg}
\ee
where the zeroth-order coefficient $b_0$ has been put equal to $1$ by
a suitable normalization of the data.

The first order approximant for the inverse velocity reads \cite{Gluzmanetal01}
\be
F_1^{*}(t)=\left( 1-{\frac{b_1}{s_1}}t\right) ^{-s_1} ~. \label{vnsnks}
\ee
The second order approximant is
\be
F_2^{*}(t)=1+b_1t\left( 1-{\frac{b_2}{b_1s_2}}t\right) ^{-s_2} ~.
\label{bvbnnc}
\ee
The exponents $s_1$ and $s_2$ are control parameters that are determined
from an optimal stability criterion. We follow \cite{Gluzmanetal01}
and impose $s_{1}=s_2=s$, which is a condition of consistency
between the two approximants. $s$ is now the single control parameter,
and plays the role of the critical exponent at the critical point $t_c$.
The condition of the existence of a critical point is that both approximants
$F_1^{*}(t)$ and $F_2^{*}(t)$ of the inverse velocity should vanish at $t=t_c$.
This yields two equations determining $t_c$ and $s$, which can be 
solved numerically.

\begin{figure}
\epsfig{file=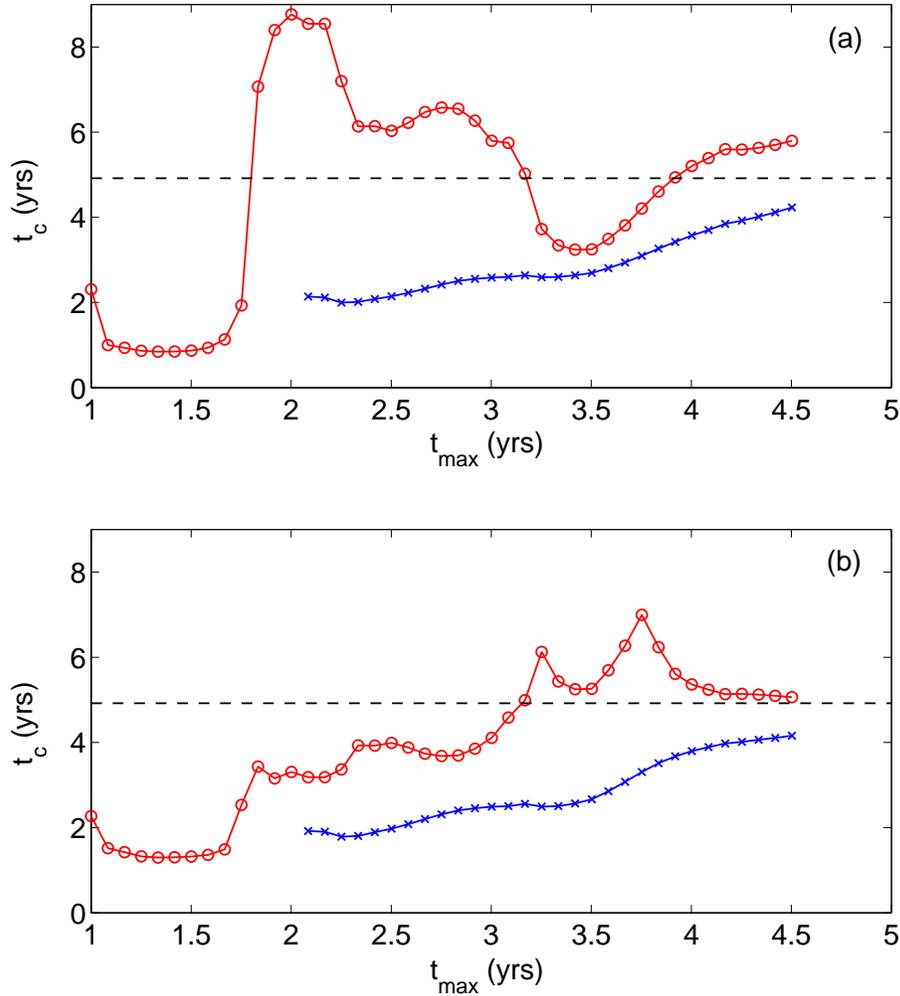,width=12cm}
\caption{Panel (a): prediction of a critical time
using a fit with a polynomial of order two in time
to the inverse of the velocity; panel (b): prediction
of the renormalization approach described in Appendix B.
In each panel, two curves are presented
corresponding to two different starting points of the data taken into 
account in the
predictions: the leftmost points correspond to the
first date taken into account in the predictions; the predictions corresponding
to the crosses $\times$ use approximately two years fewer data than 
the predictions shown
with the open circles.
The abscissa $t_{\rm max}$ is the running ``present time'', that is, the last
time of the data taken into account to issue a prediction.
The maximum realized velocity occurred at a time indicated by the
horizontal dashed line. This time is thus a proxy for the ghost-like critical
time of the landslide.
}
\label{renorm}
\end{figure}

The numerical estimates of $(t_c, s)$ depends on the time interval
over which the polynomial coefficients $b_1$ and $b_2$ are determined. Let
$t_{\rm max}$ denote the last point used in the polynomial fit.
Figure \ref{renorm} shows the numerical estimate of $t_c$ as a function of
$t_{\rm max}$. More precisely,
Figure \ref{renorm} compares the prediction of a fit using a 
polynomial of order two in time
to the inverse of the velocity (panel (a)) with the prediction
of the renormalization approach (panel (b)).

We have also fitted a power law to the data to extract
an estimate of $t_c$ as a function of $t_{\rm max}$ and find an extremely
unstable prediction where $t_c$ fluctuates wildly
ranging from two years before the end of 1987 to 25 years after 1987.
Clearly, predicting the change of regime from a power law fit of the 
acceleration
in the first phase of La Clapi\`ere is completely unreliable.
In contrast, the renormalized approximants provide a more reasonable 
stable estimate.

\newpage

\clearpage

\clearpage

\clearpage

\clearpage

\clearpage

\clearpage

\clearpage

\clearpage

\clearpage

\end{document}